\documentclass[10pt,conference]{IEEEtran}
\IEEEoverridecommandlockouts

\usepackage{flushend}
\usepackage{graphicx}
\usepackage{booktabs}
\usepackage{array}
\usepackage{xcolor}
\usepackage{listings}
\usepackage{fontawesome5}
\usepackage{url}

\definecolor{iconhue}{HTML}{0072B2}

\DeclareUrlCommand\ghpath{\urlstyle{tt}}

\usepackage[T1]{fontenc}
\usepackage[utf8]{inputenc}
\usepackage{cite}
\usepackage{orcidlink}
\usepackage{relsize}
\usepackage[varqu]{zi4}  
\usepackage{microtype}
\DisableLigatures[-]{encoding=T1,family=tt*}  
\usepackage{fontawesome5}
\usepackage{listings}
\usepackage{enumitem}
\usepackage{url}
\usepackage{hyperref}
\definecolor{ACMBlue}{RGB}{0,85,162} 
\hypersetup{
    colorlinks=true,
    citecolor=ACMBlue,
    filecolor=black,
    linkcolor=ACMBlue,
    urlcolor=ACMBlue
    }

\usepackage[labelsep=colon,font=small]{caption}

\makeatletter
\renewcommand\paragraph{\@startsection{paragraph}{4}{\z@}%
  {1.1ex \@plus 0.5ex \@minus 0.2ex}{-0.6em}%
  {\normalfont\normalsize\bfseries}}
\makeatother

\renewcommand{\lstlistingname}{\bfseries Listing}
\makeatletter
\def\fnum@lstlisting{%
  \lstlistingname
  \ifx\lst@@caption\@empty\else~\thelstlisting\normalfont\fi}%
\makeatother

\graphicspath{{figures/}}

\def\seartQueryDate{\NA}        
\def\cloneStartDate{\NA}        
\def\cloneEndDate{\NA}          

\def\nAgents{\NA}               
\def\nCandidateRepos{\NA}       
\def\nInactiveDropped{\NA}      
\def\nComparableRepos{\NA}      
\def\nErrorRepos{\NA}           
\def\nPreRelease{\NA}           
\def\nPreReleaseHistories{\NA}  

\def\nSampled{\NA}              
\def\nSampledNamed{\NA}         
\def\agreeOverall{\NA}          
\def\kappaOverall{\NA}          

\def\nHistories{\NA}            
\def\nRevisions{\NA}            
\def\nInvocations{\NA}          
\def\nInvWeak{\NA}              
\def\nInvUses{\NA}              
\def\nInvRun{\NA}               
\def\nInvPath{\NA}              
\def\nInvGhaw{\NA}              
\def\nInvUrl{\NA}               
\def\nInvEnv{\NA}               
\def\nAgenticRepos{\NA}         
\def\pctAgenticReposRecent{\NA} 
\def\pctAgenticReposComparable{\NA}
\def\nGenericRepos{\NA}         
\def\nGhawRepos{\NA}            
\def\nGhawManifests{\NA}        
\def\nAgenticReposLast{\NA}     
\def\nLatestAgentic{\NA}        

\def\wiringFirstQuarter{\NA}    
\def\wiringLastQuarter{\NA}     
\def\pctWireUsesFirst{\NA}      
\def\pctWireUsesLast{\NA}       
\def\pctWireRunFirst{\NA}       
\def\pctWireRunLast{\NA}        
\def\pctWirePathFirst{\NA}      
\def\pctWirePathLast{\NA}       
\def\pctWireGhawFirst{\NA}      
\def\pctWireGhawLast{\NA}       
\def\pctTaskReview{\NA}         
\def\pctTaskTriage{\NA}         
\def\pctTaskImplement{\NA}      
\def\pctGatedComment{\NA}       
\def\pctGatedPRTarget{\NA}      
\def\pctPermComment{\NA}        
\def\pctPinComment{\NA}         
\def\nTrigPRTarget{\NA}         
\def\nGhawSourcesLast{\NA}      
\def\pctGhawNetwork{\NA}        
\def\pctGhawSandbox{\NA}        
\def\nSameDayAbandon{\NA}       
\def\pctAbandonMonth{\NA}       
\def\pctAbandonQuarter{\NA}     
\def\pctAbandonYear{\NA}        
\def\codexReachMonths{\NA}      
\def\nCodexYear{\NA}            
\def\nRollbackPulumi{\NA}       
\def\nRollbackRlang{\NA}        

\IfFileExists{numbers.tex}{

\def\seartQueryDate{3 October 2026}
\def\nCandidateRepos{196{,}938}

\def\nComparableRepos{172{,}607}

\def\nInactiveDropped{143{,}943}

\def\nAgents{12}

\def\nRollbackPulumi{33}
\def\nRollbackRlang{32}
}{}
\IfFileExists{numbers-dataset.tex}{

\def\cloneEndDate{5 October 2026}
\def\cloneStartDate{3 October 2026}
\def\codexReachMonths{8}
\def\firstInvocationDate{16 July 2023}

\def\nAgenticRepos{6{,}345}

\def\nAgenticReposLast{4{,}598}

\def\nAgenticRevisions{125{,}266}
\def\nAgenticRevisionsLast{9{,}265}
\def\nCodexYear{21}
\def\nCommits{104{,}545}
\def\nCommitsLast{6{,}923}
\def\nErrorRepos{681}

\def\nGenericRepos{417}
\def\nGhawManifests{36{,}660}
\def\nGhawRepos{506}
\def\nGhawReposLast{377}
\def\nGhawSourcesLast{1{,}120}
\def\nHistories{15{,}928}

\def\nHistoriesLast{9{,}942}

\def\nInvEnv{24{,}963}
\def\nInvGhaw{36{,}657}
\def\nInvGhawLast{2{,}153}

\def\nInvPath{8{,}805}
\def\nInvPathLast{992}
\def\nInvRun{64{,}822}
\def\nInvRunLast{3{,}421}
\def\nInvUrl{1{,}969}
\def\nInvUses{69{,}104}
\def\nInvUsesLast{5{,}050}
\def\nInvWeak{26{,}932}
\def\nInvocations{205{,}832}

\def\nInvocationsLast{12{,}944}
\def\nLatestAgentic{9{,}265}

\def\nPreRelease{3}
\def\nPreReleaseHistories{3}

\def\nRevisions{159{,}878}

\def\nRevisionsLast{9{,}942}
\def\nSameDayAbandon{174}

\def\nTrigPRTarget{327}

\def\pctAbandonMonth{10}
\def\pctAbandonQuarter{20}
\def\pctAbandonYear{44}

\def\pctAgenticReposComparable{2.6}
\def\pctAgenticReposLast{72.5}
\def\pctAgenticReposRecent{7.9}
\def\pctAgenticRevisionsLast{7.4}
\def\pctCommitsLast{6.6}

\def\pctGatedComment{85}

\def\pctGatedPRTarget{55}

\def\pctGhawNetwork{50.8}

\def\pctGhawReposLast{74.5}
\def\pctGhawSandbox{2.7}

\def\pctHistoriesLast{62.4}
\def\pctInvGhawLast{5.9}
\def\pctInvPathLast{11.3}
\def\pctInvRunLast{5.3}
\def\pctInvUsesLast{7.3}
\def\pctInvocationsLast{6.3}
\def\pctPermComment{19}

\def\pctPinComment{24}

\def\pctRevisionsLast{6.2}

\def\pctTaskImplement{5.7}
\def\pctTaskReview{38.4}
\def\pctTaskTriage{34.4}

\def\pctWireGhawFirst{0.4}
\def\pctWireGhawLast{26.6}
\def\pctWirePathFirst{18.4}
\def\pctWirePathLast{3.7}
\def\pctWireRunFirst{7.3}
\def\pctWireRunLast{27.6}
\def\pctWireUsesFirst{73.9}
\def\pctWireUsesLast{42.1}

\def\wiringFirstQuarter{2025 Q2}
\def\wiringLastQuarter{2026 Q3}
}{}
\IfFileExists{numbers-validation.tex}{

\def\agreeOverall{97.7}
\def\kappaOverall{0.96}

\def\nSampled{576}
\def\nSampledNamed{384}

}{}

\newcommand{\dataset}{GHAgentCI}
\newcommand{\code}[1]{\textsf{#1}}
\newcommand{\awe}{\code{cai-miner}}
\newcommand{\cofee}{\code{cofee}}
\newcommand{\eg}{\emph{e.g.},\ }

\begin{document}
\bstctlcite{IEEEcontrol:no_dash}

\title{\dataset: A Dataset of Coding-Agent Invocations in CI Workflow Histories}

\author{
\IEEEauthorblockN{Daniel Ogenrwot~\orcidlink{0000-0002-0133-8164},
Jorge Gonzalo Delgado Cervantes~\orcidlink{0009-0005-5155-8914},
Woojoong Kim~\orcidlink{0009-0007-8068-1035},
John Businge~\orcidlink{0000-0003-3206-7085}}
\IEEEauthorblockA{\textit{University of Nevada Las Vegas}\\
Las Vegas, USA\\
\{ogenrwot, delgaj10, kimw19\}@unlv.nevada.edu, john.businge@unlv.edu}
}

\maketitle

\begin{abstract}

Coding agents are increasingly integrated into GitHub Actions to review pull requests, triage issues and modify code. Identifying these integrations is challenging because agentic workflows often resemble ordinary CI files, with evidence appearing in action references, shell commands and workflow conventions. We present GHAgentCI, a dataset of 15K+ workflow histories from 6K+ of 196K+ mined GitHub repositories, comprising 159K+ revisions and 205K+ agent-evidence records linked to workflow triggers and declared execution constraints. Our open-source extractor, \awe{}, detects agent-related evidence using an extensible catalogue. Manual validation of 576 stratified records confirmed 96.1\% of dedicated-action matches and 98.8\% of GitHub Agentic Workflow matches, with lower rates for other evidence kinds. Exploratory analyses reveal a shift toward command-line and GitHub Agentic Workflow integrations, differences in declared safeguards, and frequent removal of detected agentic workflows. GHAgentCI supports longitudinal studies of coding-agent integration into CI. 

\end{abstract}


\section{Introduction}
\label{sec:introduction}

Coding agents are increasingly integrated into repository
automation. GitHub Actions can invoke them on pull
requests, issue comments and schedules to review
changes, triage issues and modify code. Dedicated actions
such as \code{claude-code-action}~\cite{claude_code_action},
command-line tools and GitHub Agentic Workflows
(gh-aw)~\cite{gh_aw}
support this automation. Unlike interactive agent use,
which has received research
attention~\cite{robbes2026promises,li2025aidev,AgenticFlict,How-AI-Coding,ogenrwot2025patchtrackcomprehensiveanalysischatgpts,ogenrwot:2024ase,sserunjogi2026promptqualitypullrequest},
these workflows can process external contributions
under CI-defined permissions and conditions.
Studying this practice requires examining invocation
mechanisms, triggers, declared constraints and
their evolution.

Identifying these integrations at scale is challenging. Agentic workflows often lack distinctive filenames under \code{.github/workflows/}. Evidence may appear in action references, shell commands or \code{gh-aw} compiler
metadata. Copilot's reserved setup workflow configures agents without invoking them, while credentials such as \code{ANTHROPIC\_API\_KEY} may appear without an agent invocation. Existing datasets cover agent-authored changes, agent configuration files or general workflow histories~\cite{li2025aidev,cardoen2026ghagentfiles, cardoen2024workflows}, but do not systematically link agent-related evidence to workflow triggers, declared constraints and their evolution. 

To address this gap, we present \dataset{}, a dataset of 15K+ workflow-file histories from 6K+ GitHub repositories, containing 159K+ revisions and 205K+ agent-evidence records. The dataset links step- and file-level evidence to workflow triggers, permissions and actor conditions (Section~\ref{sec:dataset}).To demonstrate its research potential, we characterise how agent integration mechanisms, declared execution constraints and workflow retention have changed over time (Section~\ref{sec:findings}).
\section{Related Work}
\label{sec:related}

Prior research examines coding-agent outputs and configurations. AIDev collects agent-authored pull requests~\cite{li2025aidev}, while Robbes \emph{et al.} study adoption through commits and pull requests~\cite{robbes2026promises,robbes2026agentic}. Other studies investigate agent context and configuration files~\cite{mohsenimofidi2026context,chatlatanagulchai2025agentreadmes,galster2026configuring,galster2026dataset}, with GHAgentFiles providing longitudinal histories~\cite{cardoen2026ghagentfiles}. Ghaleb~\cite{ghaleb2026cicd} studies agent-authored pull requests modifying CI configurations. These works examine agent instructions and outputs rather than their invocation from CI workflows. Research on GitHub Actions examines adoption and use~\cite{kinsman2021actions,decan2022actions}, migration from other CI services~\cite{golzadeh2022ci}, and effects on pull requests~\cite{wessel2023actions}. Cardoen \emph{et al.}~\cite{cardoen2024workflows} provide general GitHub Actions workflow histories without identifying coding-agent integrations. GHAW-H~\cite{valenzuela2026ghawh} tracks GitHub Agentic Workflows in 262 repositories, linking Markdown specifications to compiled lock files, but covers only gh-aw. Neither dataset systematically detects integrations across dedicated actions, shell commands and other mechanisms. \dataset{} complements these datasets with longitudinal evidence of coding-agent integrations, workflow triggers and declared safeguards. Repository and file-level metadata support linkage with
existing datasets.
\section{Detecting Agent Invocations}
\label{sec:detection}

We developed \awe{}, an open-source tool for detecting agent-related evidence in GitHub Actions workflow histories. Unlike agent configuration files, agentic workflows often have no distinctive filenames under \code{.github/workflows/} and must be identified by
their contents. Based on vendor documentation and existing workflows, we identified four integration mechanisms. First, \emph{dedicated actions}, such as \code{anthropics/claude-code-action}, invoke agents through workflow steps. Second, shell steps may execute agent \emph{command-line interfaces} (CLIs), such as \code{claude -p} or \code{codex exec}. Third, Copilot's \emph{reserved workflow file}, \code{copilot-setup steps.yml}, configures its execution environment. Fourth, GitHub Agentic Workflows (\emph{gh-aw}) compiles Markdown specifications into \code{\*.lock.yml} workflows whose headers declare the agent engine. Detecting only dedicated actions would therefore miss other integration mechanisms.

We encode these traces in a declarative TOML \emph{catalogue} (Listing~\ref{lst:catalogue}). The \code{[paths]} section specifies files to extract, including gh-aw Markdown sources, while agent sections define detection patterns for the evidence kinds in Table~\ref{tab:evidence}. We constructed the catalogue from documentation for nine agents: Claude, Codex, Copilot, Gemini, Cursor,
Aider, Devin, PR-Agent and CodeRabbit. We also included 1,122 action repositories listed in GitHub Marketplace's \code{ai-assisted} category on 8 September 2026. Actions associated with known vendors added Goose, OpenCode and OpenHands; the remaining actions form a \code{generic\_ai} group, which may include general LLM utilities. The catalogue covers \nAgents\ named agents and was frozen before mining. We retain \code{generic\_ai} matches in the dataset but exclude them from the reported adoption counts and analyses.

\begin{lstlisting}[float=t,caption={Excerpt of the frozen catalogue: the files to extract, and the patterns that identify Claude and Copilot, by evidence kind.},label={lst:catalogue}]
[paths]
github_actions    = [".github/workflows/*.yml",
                     ".github/workflows/*.yaml"]
agentic_workflows = [".github/workflows/*.md"]

[agents.claude]
uses = ["anthropics/claude-code-action",
        "anthropics/claude-code-base-action"]
run  = ["claude -p", "claude --print",
        "@anthropic-ai/claude-code", ...]
url  = ["api.anthropic.com"]
env  = ["ANTHROPIC_API_KEY", "CLAUDE_CODE_OAUTH_TOKEN"]

[agents.copilot]
path = ["copilot-setup-steps.yml",
        "copilot-setup-steps.yaml"]
job  = ["copilot-setup-steps"]
...
\end{lstlisting}

\begin{table}[t]
\centering
\caption{Evidence kinds, example patterns, evidence tiers, and the number of agent-evidence records associated with each kind. Only strong evidence classifies a workflow revision as agentic.}
\label{tab:evidence}
\footnotesize
\setlength{\tabcolsep}{3pt}
\begin{tabular}{@{}l l l l r@{}}
\toprule
\textbf{kind} & \textbf{matched in} & \textbf{example} & \textbf{tier} & \textbf{records} \\
\midrule
\code{uses} & step \code{uses:} & \code{anthropics/claude-code-action} & strong & \nInvUses \\
\code{run}  & step \code{run:}  & \code{claude -p} & strong & \nInvRun \\
\code{path} & file name & \code{copilot-setup-steps.yml} & strong & \nInvPath \\
\code{ghaw} & lock-file header & engine declared by gh-aw & strong & \nInvGhaw \\
\code{job}  & job identifier & \code{copilot-setup-steps} & weak & --\textsuperscript{*} \\
\code{url}  & step content & \code{api.anthropic.com} & weak & \nInvUrl \\
\code{env}  & \code{env:}, any scope & \code{ANTHROPIC\_API\_KEY} & weak & \nInvEnv \\
\bottomrule
\multicolumn{5}{@{}p{\columnwidth}@{}}{\textsuperscript{*}\scriptsize Reserved job
identifiers occur only together with the reserved file name, which decides.}
\end{tabular}
\end{table}

Given a local repository clone, \awe{} traverses the default branch's first-parent history, following \cofee{}~\cite{cardoen2026ghagentfiles}. It records workflow revisions with their git metadata and change types. A logical file retains its identifier across modifications and renames, but receives a new one if deleted and recreated. Each distinct YAML revision is parsed and matched against the catalogue at the job and step levels. For gh-aw, \awe{} parses compiler headers and Markdown frontmatter across compiler versions, extracting the declared engine, tools, permitted outputs, network policy and sandbox. Unparseable revisions fall back to text matching and are flagged accordingly.

Because not every match indicates an invocation, we distinguish strong evidence from weaker signals, such as credentials and inference endpoints, which may also occur in ordinary LLM-backed applications. Weak evidence alone does not classify a revision as agentic. Records retain their declaring scope: a workflow-level \code{ANTHROPIC\_API\_KEY} produces one file-level record, not one per step. Multiple matches within a step are combined into one record, retaining the strongest evidence kind. Reserved Copilot setup files remain configuration evidence, not direct invocations. Each record preserves its evidence kind and scope, distinguishing invocation
declarations from indirect configuration evidence.
\section{Building the Dataset}
\label{sec:construction}

To study coding-agent adoption in CI, we selected repositories independently of agent usage. We queried SEART~\cite{dabic2021seart} on \seartQueryDate\ for non-fork GitHub repositories with at least 300 commits and a last-commit date within the preceding year, as recorded in the SEART export. This excluded \nInactiveDropped\ inactive repositories, leaving \nCandidateRepos\ candidates. Unlike GHAgentFiles~\cite{cardoen2026ghagentfiles}, we did not require repositories to be at least one year old, which would exclude newer projects potentially adopting coding agents. Instead, we flagged the \nComparableRepos\ repositories meeting this criterion for comparisons with GHAgentFiles.

We cloned repositories between \cloneStartDate\ and \cloneEndDate\ using partial clones excluding blobs larger than 64\,KB. To avoid truncated workflow histories, \awe{} separately fetched workflow blobs exceeding this limit and recorded blob request and read counts to check extraction completeness. Repositories exceeding the initial 15-minute mining budget were retried with a one-hour budget. After retries, \nErrorRepos\ repositories remained unmined, mostly because they were deleted or private. These and four deliberately excluded gh-aw development repositories remain in the dataset with their status. Finally, we examined matches predating LLM-based coding agents. Some Markdown frontmatter keys matched the gh-aw detector spuriously. Following GHAgentFiles~\cite{cardoen2026ghagentfiles}, we used ChatGPT's release on 30 November 2022 as a conservative cutoff, excluding \nPreRelease\ evidence records from \nPreReleaseHistories\ histories. The earliest retained agentic workflows include PR-Agent and CodeRabbit,
used for pull request reviews in 2023.
\section{Validation}
\label{sec:validation}

We sampled \nSampled\ records across strata defined by the deciding evidence kind. The \nSampledNamed\ named-agent records were allocated proportionally across \code{uses}, \code{run}, \code{path}, gh-aw and parse-failed strata, with at least 30 each, targeting a $\pm$5\% margin at 95\% confidence. Weak-only and \code{generic\_ai} evidence received 96 records each. Within each stratum, we used systematic sampling with a random start, ordered by agent, and selected at most two records per repository.

For each record, two authors independently inspected the workflow at its commit, without seeing the detector's classification. A \emph{yes} label required evidence of a coding-agent invocation or, for reserved Copilot setup files, valid agent-specific configuration. Merely naming or installing an agent was insufficient. Weak-only matches required additional evidence of agent execution, while ambiguous cases were labeled maybe. Only \emph{yes} labels counted as confirmed. The authors agreed on \agreeOverall\% of records ($\kappa = \kappaOverall$) and resolved disagreements through discussion. Table~\ref{tab:validation} reports the adjudicated labels and confirmation rates by evidence kind.

\begin{table}[t]
\centering
\caption{Manual validation by evidence kind.
Strict rates treat \emph{maybe} as incorrect.
Intervals are 95\% Wilson confidence intervals.}
\label{tab:validation}
\footnotesize
\setlength{\tabcolsep}{3pt}
\begin{tabular}{lrrrrrl}
\toprule
\textbf{Kind} & \textbf{$n$} & \textbf{Yes} & \textbf{Maybe} & \textbf{No} & \textbf{Confirmed (\%)} & \textbf{95\% CI} \\
\midrule
\code{gh-aw}        & 80  & 79 & 0  & 1  & 98.8 & [93.3,99.8] \\
\code{uses}         & 128 & 123& 3  & 2  & 96.1 & [91.2,98.3] \\
\code{run}          & 116 & 95 & 4  & 17 & 81.9 & [73.9,87.8] \\
\code{path}         & 30  & 26 & 0  & 4  & 86.7 & [70.3,94.7] \\
\code{parse-failed} & 30  & 14 & 1  & 15 & 46.7 & [30.2,63.9] \\
\code{weak-only}    & 96  & 41 & 19 & 36 & 42.7 & [33.3,52.7] \\
\code{generic AI}   & 96  & 49 & 12 & 35 & 51.0 & [41.2,60.8] \\
\midrule
Total        & 576 & 427& 39 & 110& --   & -- \\
\bottomrule
\end{tabular}
\end{table}

Dedicated actions and gh-aw yielded the highest confirmation rates (96.1\% and 98.8\%), followed by shell commands (81.9\%), some of which matched agent names or installation commands without execution. Parse-failed records reached 46.7\%, weak-only evidence 42.7\%, and generic AI actions 51.0\%. The latter two groups are retained separately rather than counted as confirmed invocations. To examine missed invocations, we re-mined 772 randomly selected repositories containing workflows and inspected 200 unflagged workflows selected using broad keywords (e.g., \texttt{llm}, \texttt{agent}). Eleven (5.5\%; 95\% Wilson CI: 3.1--9.6\%) contained an agent invocation. This is the missed-invocation rate within the keyword-selected sample, not overall recall.
\section{The \dataset{} Dataset}
\label{sec:dataset}

\dataset{} comprises four compressed CSV tables (Table~\ref{tab:dataset}) and an archive of distinct workflow revisions stored by SHA-256 content hash. The tables are keyed by repository, file identifier and commit hash. The repository table includes agentic and non-agentic repositories, providing an adoption denominator. The revision table tracks each change to a file that is agentic at some point, from creation to deletion, and retains the git metadata of GHAgentFiles~\cite{cardoen2026ghagentfiles} to support linking the datasets. Each steps-table row records agent-related evidence at the \emph{step} level (job and step) or \emph{file} level (reserved filename, workflow-level credential or gh-aw header). A field-level schema accompanies the dataset, which is released with \awe{}.

\begin{table}[t]
\centering
\caption{Content of the dataset}
\label{tab:dataset}
\footnotesize
\begin{tabular}{@{}>{\raggedright\arraybackslash}p{2.9cm} p{5.2cm}@{}}
\toprule
\textbf{file} & \textbf{description} \\
\midrule
\code{revisions.csv.gz} & Compressed CSV file with one row per touch of a
workflow file in an agentic history. \\
\code{steps.csv.gz} & Compressed CSV file with one row per agent-evidence record per revision. \\
\code{ghaw\_manifests.csv.gz} & Compressed CSV file with one row per revision of a gh-aw artifact: engine, compiler and schema versions, model, tools, MCP tools, safe outputs, secrets, sandboxing, network policy and permissions. \\
\code{repositories.csv.gz} & Compressed CSV file with one row per mined
repository: commit count, workflow and agentic file counts, agents used, first and last agentic dates, blob accounting and status. \\
\code{content.tar.gz} & Compressed archive of every distinct workflow
revision. \\
\bottomrule
\end{tabular}
\end{table}

Table~\ref{tab:size} summarises \dataset{}. Histories span \firstInvocationDate\ to \cloneEndDate\ and include revisions before
agent introduction and after removal. \nInvWeak\ records contain weak evidence only. The share of repositories with detected agentic workflow
histories is \pctAgenticReposRecent\%  among repositories younger than one year, compared with \pctAgenticReposComparable\% among older repositories. Another \nGenericRepos\ repositories contain only \code{generic\_ai} actions and are kept separate. The dataset also includes \nGhawManifests\ gh-aw artifact revisions from \nGhawRepos\ repositories. 

\begin{table}[t]
\centering
\caption{Characteristics of the dataset. The \% indicates the proportion of elements in the last snapshot w.r.t.\ all snapshots.}
\label{tab:size}
\footnotesize
\begin{tabular}{@{}l r r@{}}
\toprule
\textbf{characteristic} & \textbf{all snapshots} & \textbf{last snapshot} \\
\midrule
\# repositories & \nAgenticRepos & \nAgenticReposLast\ (\pctAgenticReposLast\%) \\
\quad with gh-aw workflows & \nGhawRepos & \nGhawReposLast\ (\pctGhawReposLast\%) \\
\# histories & \nHistories & \nHistoriesLast\ (\pctHistoriesLast\%) \\
\# commits & \nCommits & \nCommitsLast\ (\pctCommitsLast\%) \\
\# workflow revisions & \nRevisions & \nRevisionsLast\ (\pctRevisionsLast\%) \\
\quad agentic & \nAgenticRevisions & \nAgenticRevisionsLast\ (\pctAgenticRevisionsLast\%) \\
\# agent records & \nInvocations & \nInvocationsLast\ (\pctInvocationsLast\%) \\
\quad via \code{uses} & \nInvUses & \nInvUsesLast\ (\pctInvUsesLast\%) \\
\quad via \code{run} & \nInvRun & \nInvRunLast\ (\pctInvRunLast\%) \\
\quad via \code{path} & \nInvPath & \nInvPathLast\ (\pctInvPathLast\%) \\
\quad via \code{ghaw} & \nInvGhaw & \nInvGhawLast\ (\pctInvGhawLast\%) \\
\bottomrule
\end{tabular}
\end{table}
\section{Exploratory Empirical Analysis}
\label{sec:findings}

We examine three questions using named agents with
strong evidence only.

\noindent \textit{\textbf{RQ1: How have coding-agent integration mechanisms in CI evolved?}} 

\begin{figure}[t]
\centering
\includegraphics[scale=0.85]{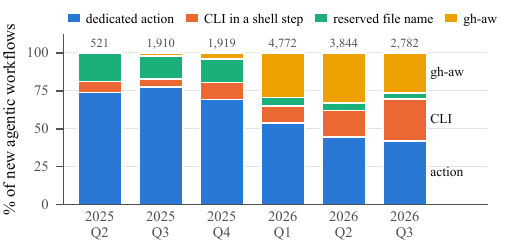}
\caption{Integration mechanisms detected in newly
agentic workflows by quarter; numbers above the bars
indicate workflow counts. Quarters that ended after
cloning started are omitted.}
\label{fig:wiring}
\end{figure}

Fig.~\ref{fig:wiring} shows how newly detected agentic workflows are integrated into CI. In \wiringFirstQuarter, \pctWireUsesFirst\% used dedicated vendor actions, compared with \pctWireUsesLast\% in \wiringLastQuarter. Over the same period, gh-aw increased from \pctWireGhawFirst\% to \pctWireGhawLast\%, and CLI invocations from \pctWireRunFirst\% to \pctWireRunLast\%, while reserved Copilot setup files declined from \pctWirePathFirst\% to \pctWirePathLast\%. At the latest repository snapshot, the most common mechanism varies by agent: dedicated actions account for 77\% of Claude workflows, CLI calls for 51\% of Codex workflows, and gh-aw for 48\% of Copilot workflows. These findings show that identifying agentic workflows solely through dedicated action references misses a growing share of them.

\noindent \textit{\textbf{RQ2: How do workflow triggers relate to execution constraints?}}

\begin{figure}[t]
\centering
\includegraphics[scale=0.85]{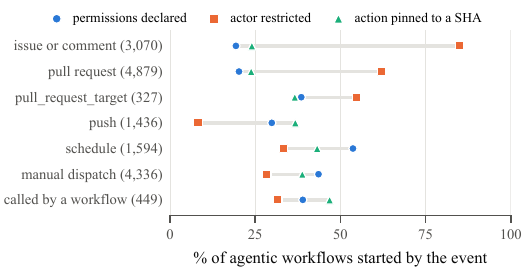}
\caption{Declared safeguards by trigger at the latest commit. Counts are workflows; those with multiple triggers appear in multiple groups. SHA pinning applies to dedicated actions.}
\label{fig:constraints}
\end{figure}

Fig.~\ref{fig:constraints} examines declared safeguards across \nLatestAgentic\ agentic workflows. Inferred tasks are mainly review (\pctTaskReview\%) and triage (\pctTaskTriage\%), rather than implementation (\pctTaskImplement\%). Among issue- or comment-triggered workflows, \pctGatedComment\% restrict triggering actors, but only \pctPermComment\% explicitly declare token permissions and \pctPinComment\% of dedicated actions are SHA-pinned. Scheduled and manually triggered workflows show the opposite pattern. Of \nTrigPRTarget\ workflows using \code{pull\_request\_target}, only \pctGatedPRTarget\% restrict triggering actors, despite the potential exposure of repository secrets to fork-triggered workflows. Among \nGhawSourcesLast\ gh-aw sources, \pctGhawNetwork\% declare custom network policies and \pctGhawSandbox\% explicit sandboxing.

\noindent \textit{\textbf{RQ3: How long do repositories retain agentic workflows?}}

\begin{figure}[t]
\centering
\includegraphics[scale=0.85]{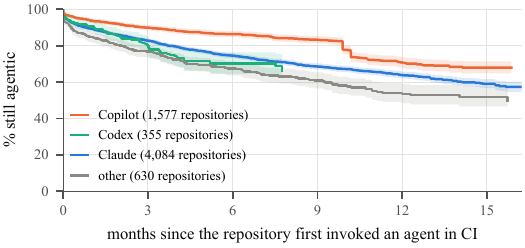}
\caption{Repository retention of detected agentic workflows by first adopted agent (Kaplan--Meier; 95\% confidence bands). Simultaneous adopters count in multiple groups; curves stop when fewer than 50 repositories remain under observation.}
\label{fig:abandonment}
\end{figure}

A repository remains agentic while any detected agentic workflow is present, regardless of workflow replacement. Fig.~\ref{fig:abandonment} shows Kaplan--Meier retention estimates: \pctAbandonMonth\% lose detected agentic workflows within one month, \pctAbandonQuarter\% within three months, and \pctAbandonYear\% within a year. Another \nSameDayAbandon\ repositories removed workflows on the adoption day. Copilot shows the longest retention; Codex's curve ends after approximately \codexReachMonths\ months, with only \nCodexYear\ repositories observed for a full year. Removals cluster across related repositories: \nRollbackPulumi\ Pulumi repositories removed Copilot setup workflows together, while \nRollbackRlang\ R-package repositories (\eg \code{r-dbi}, \code{r-lib}, \code{tidyverse}) stopped on 13 September 2026.

\paragraph{Research opportunities}
The preliminary analysis presented in this section only scratches the surface. \dataset{} supports studies of coding-agent governance and adoption over time. Linking it with GHAgentFiles~\cite{cardoen2026ghagentfiles} could compare context-file restrictions with CI safeguards, while AIDev~\cite{li2025aidev} could support comparisons of pull request review outcomes across CI-invoked and interactively used agents, where invocation origins can be established. 

\section{Threats to Validity}
\label{sec:threats}

SEART-based sampling excludes small and inactive repositories, which may adopt agents differently. Rewritten git histories may affect commit dates, while default-branch first-parent traversal misses workflows confined to other branches
~\cite{kalliamvakou2016perils}.

Like GHAgentFiles~\cite{cardoen2026ghagentfiles}, our detection catalogue relies on vendor documentation that may be incomplete or change over time, potentially biasing historical trends. \awe{} may miss invocations through unrecognised wrappers, composite actions or external reusable workflows. Some \code{run} patterns match agent names or installation commands without execution, while unparseable revisions use less precise text matching. Validation identifies false positives and some missed cases but cannot establish overall recall (Section~\ref{sec:validation}). Detected agent evidence reflects workflow declarations, not observed executions. Moving agents to external reusable workflows may appear as abandonment, while organisation-wide changes create dependencies between repository histories. Inferred \code{tasks} labels and actor-restriction flags support exploratory analysis, not ground truth or definitive security assessments.

\section{Conclusion}
\label{sec:conclusion}

We presented \dataset, a dataset of \nHistories\ agentic workflow histories spanning \nRevisions\ revisions and \nInvocations\ agent-evidence records across \nAgenticRepos\ GitHub repositories. We also provide \awe{}, an open-source extractor with an extensible catalogue for detecting agent integrations in CI. Unlike datasets focused on agent-produced artifacts or configuration files, \dataset{} captures how coding agents are integrated into CI, under which declared constraints, and how these integrations evolve. Together, the dataset and extractor support longitudinal studies of coding-agent adoption in software delivery pipelines. Future releases could extend the detection catalogue and resolve external reusable workflows, improving coverage of currently unrecognized integrations.

\section*{Data Availability}
\dataset{} and its replication package are archived on Zenodo at \url{https://doi.org/10.5281/zenodo.24088832}. The project website at
\url{https://unlv-evol.github.io/cai-miner/} provides an interactive explorer, dataset documentation and executable notebooks. The extraction tool and replication scripts are available at \url{https://github.com/unlv-evol/cai-miner}. The tool is also distributed through PyPI at \url{https://pypi.org/project/cai-miner/} and can be installed using \code{pip install cai-miner}.

\label{sec:endofcontent}

\IEEEtriggeratref{12}
\bibliographystyle{IEEEtran}
\bibliography{refs}

\end{document}